\documentclass[preprint,5p,times,authoryear,twocolumn]{elsarticle}

\usepackage{overpic}
\usepackage{graphicx}
\usepackage[english]{babel}
\usepackage{lineno}
\usepackage[ansinew]{inputenc}

\newcommand{\gca}{Geochimica et Cosmochimica Acta}

\def\mum{$\mu$m}

\journal{\gca}

\begin{document}

\begin{frontmatter}

\title{Experimental investigation of the nebular formation of chondrule rims and the formation of chondrite parent bodies}
\author{E. Beitz$^1$} \ead{e.beitz@tu-bs.de}
\author{J. Blum$^1$}
\author{R. Mathieu$^2$}
\author{A. Pack$^2$}
\author{D. C. Hezel$^{3,4}$}
\address{$^1$Institute for Geophysics and extraterrestrial Physics, Technical University Braunschweig,\\Mendelssohnstr. 3, D-38106 Braunschweig, Germany\\$^2$ Geoscience Center, University G\"ottingen \\Goldschmidtstr. 1, D–37077 G\"ottingen, Germany \\ $^3$ Natural History Museum, Department of Mineralogy, Cromwell Road, SW7 5BD, London, UK \\ $^4$ now at: University of Cologne, Department of Geology and Mineralogy\\Z\"ulpicher Str. 49b, D-50674 Cologne, Germany}

\begin{abstract}

We developed an experimental setup to test the hypothesis that accretionary rims around chondrules formed in the solar nebula by accretion of dust on the surfaces of hot chondrules. Our experimental method allows us to form dust rims around chondrule analogs while levitated in an inert-gas flow. We used micrometer-sized powdered San Carlos olivine to accrete individual dust particles onto the chondrule analogs at room temperature (20$^{\circ}$C) and at 1100$^{\circ}$C. The resulting dust rims were analyzed by means of two different techniques: non-destructive micro computer tomography, and scanning electron microscopy. Both methods give very similar results for the dust rim structure and a mean dust rim porosity of 60 percent for the hot coated samples, demonstrating that both methods are equally well suited for sample analysis. The chondrule analog's bulk composition has no measurable impact on the accretion efficiency of the dust. We measured the chemical composition of chondrule analog and dust rim to check whether elemental exchange between the two components occurred. Such a reaction zone was not found; thus, we can experimentally confirm the sharp border between chondrules and dust rims described in the literature. We adopted a simple model to derive the degree of post-accretionary compaction for different carbonaceous chondrites. Moreover, we measured the rim porosity of a fragment of Murchison meteorite, analyzed it with micro-CT and found rim porosities with this technique that are comparable to those described in the literature.

\end{abstract}


\end{frontmatter}

\section{Introduction}

Chondrites are the most common class of meteorites and make up $\sim80\,\rm vol. \%$ of the meteorites in our collections \citep[][e.g.]{bischoff_geiger:1995}. Chondrites are composed of 0-80 vol.\% chondrules, 0-100 vol.\% matrix material, 0-70 vol.\% opaque phases and 0-3 vol.\% Ca,Al-rich inclusions (\citet{Brearley&Jones:1998}; \citet{zanda:2004} ; \citet{HezelEtal:2008}; \citet{McSween:1977}). Age determination by \citet{AmelinEtal:2002} and \citet{Kurahashi2008} show that chondrules are formed in a time interval around $2.5 \pm 1.2$ million years after the first solids in the solar nebula, the Ca- and Al-rich inclusions (CAIs). Thus, chondrules were witnesses to the agglomeration phase of planet formation. Chondrules are up to several mm in diameter, often spherical and can have dust rims around them \citep{MetzlerEtal:1992} -- the latter are the focus of this study. Detailed laboratory studies provide important clues on the formation and evolution of chondrules and their rims.
\citet{Allen1980} proposed that chondrules, which had formed by some yet unknown energetic process, were freely floating in the solar nebula for some time before they were incorporated into planetesimals. During this period, the chondrules interacted with the gas and dust grains of the solar nebula. In the anticipated low-velocity collisions between chondrules and the micrometer-sized dust grains, the latter sticks to the surface of the chondrules \citep{BlumSchraepler:2004} and forms a dusty rim \citep{BeitzEtal:2011b}. There are several  competing suggestions to how long it takes to form an accretionary dust rim around a chondrule: (i) \citet{Kring1989PhDT} conclude that in regions of high dust densities accretion occurred on time scales in the order of several minutes. High dust densities are in agreement with the shock wave model as the source for the high temperature event in which chondrules formed \citep{DeschConnolly:2002}. In this model the starting temperature is about 300 K,  peak temperatures during the shock are between 1568$^{\circ}$C and 1727$^{\circ}$C, depending on chondrule number density, and after the propagation of the shock wave the temperature falls back to $>870^{\circ}$C. According to this scenario, a higher dust density is present when the ambient temperature falls below the condensation temperatures of the evaporated constituents (\citet{Lodders2003}; \citet{Ebel2000}; \citet{AlexanderEtal:2008}) and this is also consistent with a faster accretion process. (ii) \citet{Cuzzi2004} and \citet{Carballido2011876} calculate a time span of 10 to1000 years for the accretion of dust, as a result of lower dust densities in their simulations.

Here, we test the first hypothesis that chondrule rims accreted in a very fast process. We designed an experimental setup in which dust rims around artificial chondrules are produced under reducing conditions. \citet{Kring1991} and \citet{Krot:1995} constrained the temperature for rim accretion. They studied different types of chondrule rims and concluded rim accretion happened at elevated temperatures. \citet{DeschConnolly:2002} calculated a temperature in the post-shock region of $>870^{\circ}$C and slow cooling rates of $0-35\mathrm{K\, h^{-1}}$ at 1100$^{\circ}$C after the shock. Therefore, we chose a constant temperature for our experiment of 1100$^{\circ}$C. Another advantage of choosing this temperature is that the chondrule is solid at 1100$^{\circ}$C \citep{Radomsky&hewins:1990} and the formation of a sharp boundary between rim and chondrule \citep{MetzlerEtal:1992} can be tested. In addition, we performed the same experiments at room temperature to constrain whether or not the temperature influences the morphology and shape of the rim.


In the first part of this study, three experiments were preformed in which a dust rim was added to the surface of chondrule analogs. Their individual porosities were measured using mirco-CT (computer aided tomography) and  BSE-images (back scattered electron).
In the second part, we use a simple model to derive the degree of compaction for different CV chondrites. In the model, we assume that the porosity of the accreted dust rim in our experiment at 1100$^{\circ}$C is comparable to the matrix porosity. For the chondrule rim accreted at room temperature we choose a fragment of the Murchison CM2 chondrite to be analyzed by micro-CT and be compared to the porosities of the chondrule analogs' rims.

\begin{figure}[t]
    \begin{center}
        \includegraphics[width=7 cm]{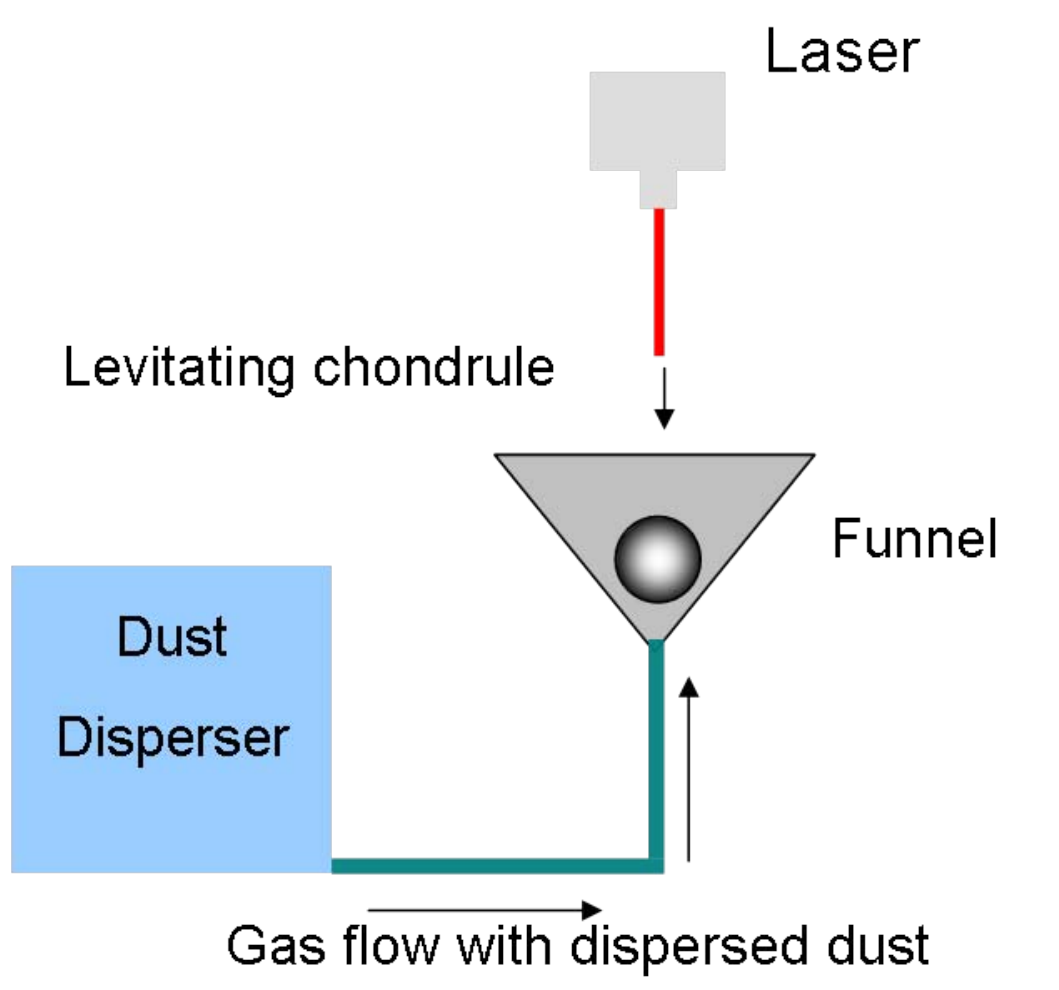}
        \caption{\label{fig:setup} Sketch of the setup at the University of G"ottingen to levitate mm-sized chondrule analogs in a vertical gas flow (for details of the apparatus, see \citet{PackEtal:2010}). The dust disperser provides the gas stream with embedded single $\rm \mu m$-sized dust grains, while the laser heats the levitated chondrule to temperatures up to and above its liquidus.}
    \end{center}
\end{figure}

\section{Levitation Experiments}
\subsection*{Experimental Setup}\label{sec:setup_experiment}
To experimentally test the hypothesis that accretionary rims were formed around free-floating chondrules, we used an improved version of the levitation apparatus at the University of G\"ottingen that is described in \citet{PackEtal:2010}. The experimental setup consists of a funnel in which the chondrule analog with a diameter of up to 2 mm can be levitated in a vertical gas flow. A 50 W CO$_2$ laser (manufacturer SYNRAD) is mounted such that the chondrule analog can be heated from the top, while being levitated in the funnel \citep{mathieuandpack:2011}. A commercial dust disperser (RGB 1000, Palas GmbH) is attached to the setup and provides a gas flow enriched with homogeneously dispersed single, micrometer-sized dust grains through the funnel. The schematics of the setup are shown in Figure \ref{fig:setup}. This levitation technique has already been used in earlier experiments to form highly porous dust rims around millimeter-sized glass beads at room temperature \citep{BeitzEtal:2011b}. The dust particles in the previous study were monodisperse, spherical $\mathrm{SiO_2}$ grains with 1.5 $\mu$m diameter. \citet{BeitzEtal:2011b} measured a rim porosity of around 82 vol.\%, which is very close to the BPCA (Ballistic Particle-Cluster Agglomeration) limit. \citet{BeitzEtal:2011b} excluded collisions of the dust-coated chondrule analog with the funnel wall because such collisions would have significantly decreased the porosity. \citet{GundlachEtal:2011} measured the dispersion efficiency of the dust disperser and found for the monodisperse $\rm SiO_2$ grains that more than 75 \% of the dust particles are being perfectly de-agglomerated into monomers, 15 \% of the particles are dimers, and 10 \% of the particles are trimers or higher polymers. We conclude from these measurements that in our study mostly monomer dust grains collide with the chondrule analog and that no rim compaction due to collisions with the funnel wall occurs.

\subsection*{Experimental Samples}
We produced a total of 3 samples. The chondrule analogs had either a forsterite or a spinel composition and were produced by mixing reagent grade oxides ($\mathrm{SiO_2}$, $\mathrm{Al_2O_3}$, and MgO) and carbonates ($\mathrm{CaCO_3}$) in the proportions given by \citet{MathieuEtal:2011}. The material was dryly ground to a fine powder in an agate mortar. Two starting compositions (from \citet{mathieu_phd}) were used for the experiments: i) Fo3 composed of 47.14 wt\% $\mathrm{SiO_2}$, 20.23 wt\% $\mathrm{Al_2O_3}$, 17.79 wt\% CaO, and 14.61 wt\% MgO (later Fo3-chondrule analog) and ii) Sp1 composed of 43.73 wt\% $\mathrm{SiO_2}$, 25.55 wt\% $\mathrm{Al_2O_3}$, 16.40 wt\% CaO, and 14.12 wt\% MgO (later Sp1-chondrule analog). Their respective liquidus temperatures are 1320$^{\circ}$C and 1405$^{\circ}$C and their solidus temperatures are 1274$^{\circ}$C and 1319$^{\circ}$C \citep{presnall1978liquidus}. The powder was pre-fused inside a mm-sized carbon cup using a laser \citep{PackEtal:2010}. The resulting irregular-shaped, crystallized silicate piece was then introduced in the gas flow inside the funnel and subsequently heated above the liquidus to form a glassy, spherical chondrule analog.

Of the 3 samples, one spinel-chondrule analog (hot Sp1) and one olivine chondrule analog (Fo3) were used for the dust accretion experiments at 1100 $^{\circ}$C. The second spinel-chondrule analog (cold Sp1) was coated at room temperature. The Fo3-chondrule analog had a diameter of 1.5 mm and was analyzed by means of micro-CT at the Natural History Museum in London and by means of BSE-images at the TU Braunschweig. The hot Sp1-chondrule analog had a diameter of 1.6 mm and was analyzed by means of standard SEM technique and a micro-CT at the TU Braunschweig. The cold Sp1 chondrule analog was only analyzed by SEM at the TU Braunschweig. All produced chondrule analogs posses bulk densities of $\sim 2.7\, \mathrm{g\,cm^{-3}}$. The densities were calculated using the commercial software Magma Density Calculator by Grabbo Soft.
\begin{figure}[t]
    \begin{center}
        \includegraphics[width=9 cm]{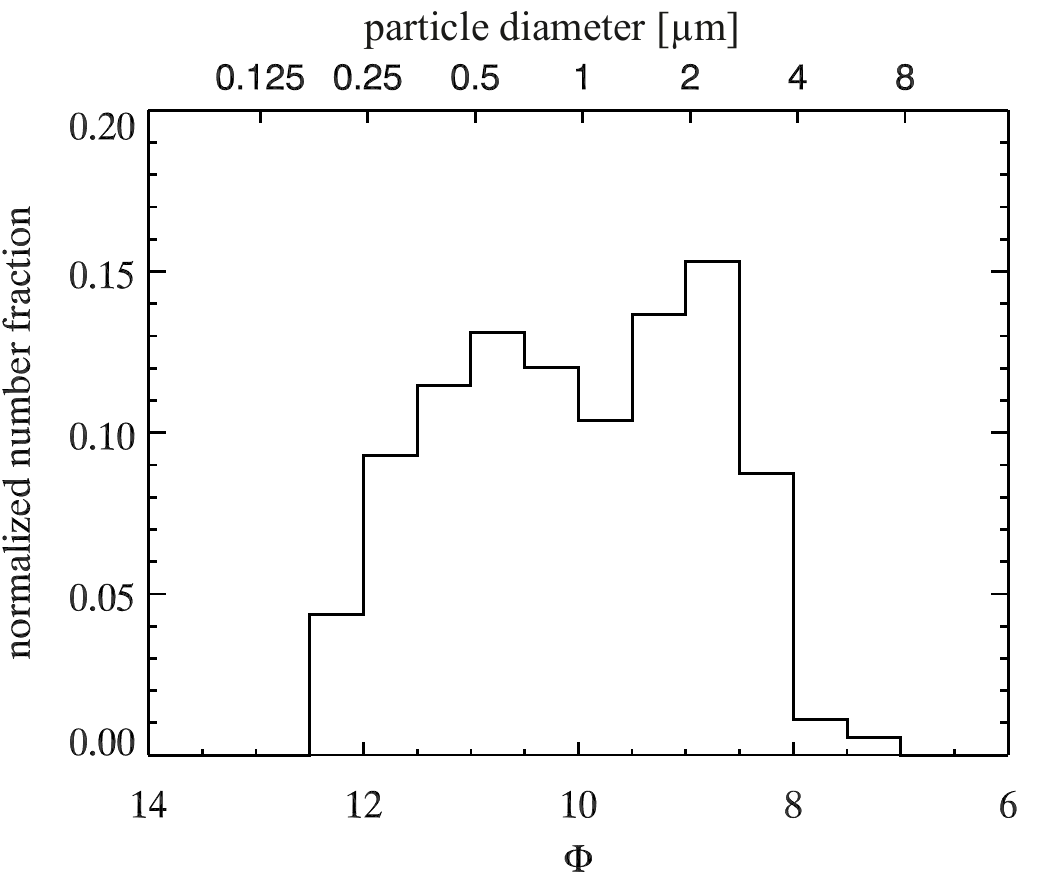}
        \caption{\label{fig:size_dis_olivine} Size distribution of the olivine dust grains that stuck to the chondrule analog's surface. Plotted is the normalized number fraction versus the diameter of the dust grains. The parameter $\phi$ is a measure for the grain diameter, i.e. $\rm\phi= -\log_{2}(grain\ diameter\ in\ mm)$.}
    \end{center}
\end{figure}

To produce the accretionary dust rims, the chondrule analogs were placed in the gas flow inside the funnel. For the rim aggregation at $1100^\circ$C we used a $90\%-10\%$ argon-hydrogen mixture to aerodynamically levitate the chondrule analogs. The hydrogen was added to prevent oxidization of the Fe content in the olivine dust. The cold Sp1 chondrule analog was levitated using air. To produce the dust rims around the chondrule analogs and to distinguish between chondrule analog and dust rim after the experiment, San Carlos olivine (San Carlos, Arizona, USA) was chosen as the aggregating dust grains. In San Carlos olivine we measured an Fe abundance of $\sim 6$ wt.\% using the EDX of the SEM. The Fe content is used to trace the boundary between the dust rim and the Fe-free chondrule analogs. The individual olivine dust grains have irregular shapes, an initial size range between 0.1 and 10 \mum, and a density of $\sim 3.2\, \mathrm{g\,cm^{-3}}$. The size distribution of the olivine grains stuck to the surface of the chondrule analogs can be determined from BSE-images of the cold Sp1 chondrule analog. For this, we binarized a BSE-image and measured the cross-sectional areas of all visible dust grains. From these measurements we calculated dust grain diameters assuming a spherical shape of the dust grains. The particle size distribution is shown in Figure \ref{fig:size_dis_olivine}, in which the normalized number fraction is plotted versus the grain diameter. We found that the experimental dust grain size distribution is in agreement to the measured dust grain sizes of fine grained rims of CM chondrites \citep{Ashworth1977}, but is truncated at grain sizes of 4 \mum, due to the low sticking efficiency of larger dust grains.

In both experimental runs at 1100 $^\circ$C, we heated the chondrule analogs to $1100^\circ$C ($\pm 30^\circ$C, \citep{mathieuandpack:2011}), which takes a few seconds, and then introduced the olivine dust in the gas flow for a duration of 120 s. The temperature is controlled via the laser energy using the LabView program \citep{mathieuandpack:2011}. After the coating time, the laser was abruptly switched off and the floating chondrule cooled down to room temperature in less then ten seconds \citep{nagashima2006reproduction}. The cooled chondrule analogs with the formed dust rims were carefully extracted using tweezers. The hot Sp1-chondrule analog was embedded in epoxy resin to be analyzed with the SEM technique first and then with the mirco-CT. The Fo3-chondrule analog was carefully stored in cellular plastic and first analyzed with the micro-CT at the Natural History Museum London and then embedded in epoxy resin to be analyzed with the SEM. Details on all measurements are summarized in Table \ref{table_method}.

\section{Analytical Methods}
We used SEM and micro-CT to analyze the dust rims of Fo3 and hot Sp1. The rim of the cold Sp1 chondrule analog was only analyzed with the SEM. Details on each sample analysis are summarized in Table \ref{table_method}. In addition, we used micro-CT to measure chondrule rim and matrix porosity of a fragment of Murchison (CM2).
\begin{table*}[t]
\center
    \caption{List of measurement methods and conditions for the dust coated chondrule analogs}
    \label{table_method}
\begin{tabular}{c c c c c }
\hline
     Composition & Temperature [$^\circ$C] & Embedded & Method & Resolution \\
  \hline

  Fo3 & 1100 & no & micro-CT $^\flat$ & 1.8 $\rm\frac{\mu m}{voxel}$\\
  Fo3 & 1100 & yes & SEM $^\dag$ & 0.06 $\rm\frac{\mu m}{pixel}$\\
  Sp1 & 1100 & yes & micro-CT $^\natural$ & 2.1 $\rm\frac{\mu m}{voxel}$\\
  Sp1 & 1100 & yes & SEM $^\sharp$ & 0.11 $\rm\frac{\mu m}{pixel}$\\
  Sp1 & 20 & yes & SEM  $^\dag$ & 0.06 $\rm\frac{\mu m}{pixel}$\\
  \hline
\end{tabular}\\
\footnotesize
$^\flat$ Measurement performed at the Natural History Museum, London.,\newline $^\dag$ Measurement performed at the Institut f\"ur Halbleitertechnik (TU Braunschweig). \newline $^\natural$ Measurement performed at  the Institut f\"ur Partikeltechnik (TU Braunschweig). \newline $^\sharp$ Measurement performed at the Institut f\"ur theoretische und physikalische Chemie (TU Braunschweig).
\end{table*}

            \begin{figure*}[p]
                \begin{center}
                    \includegraphics[width=\textwidth]{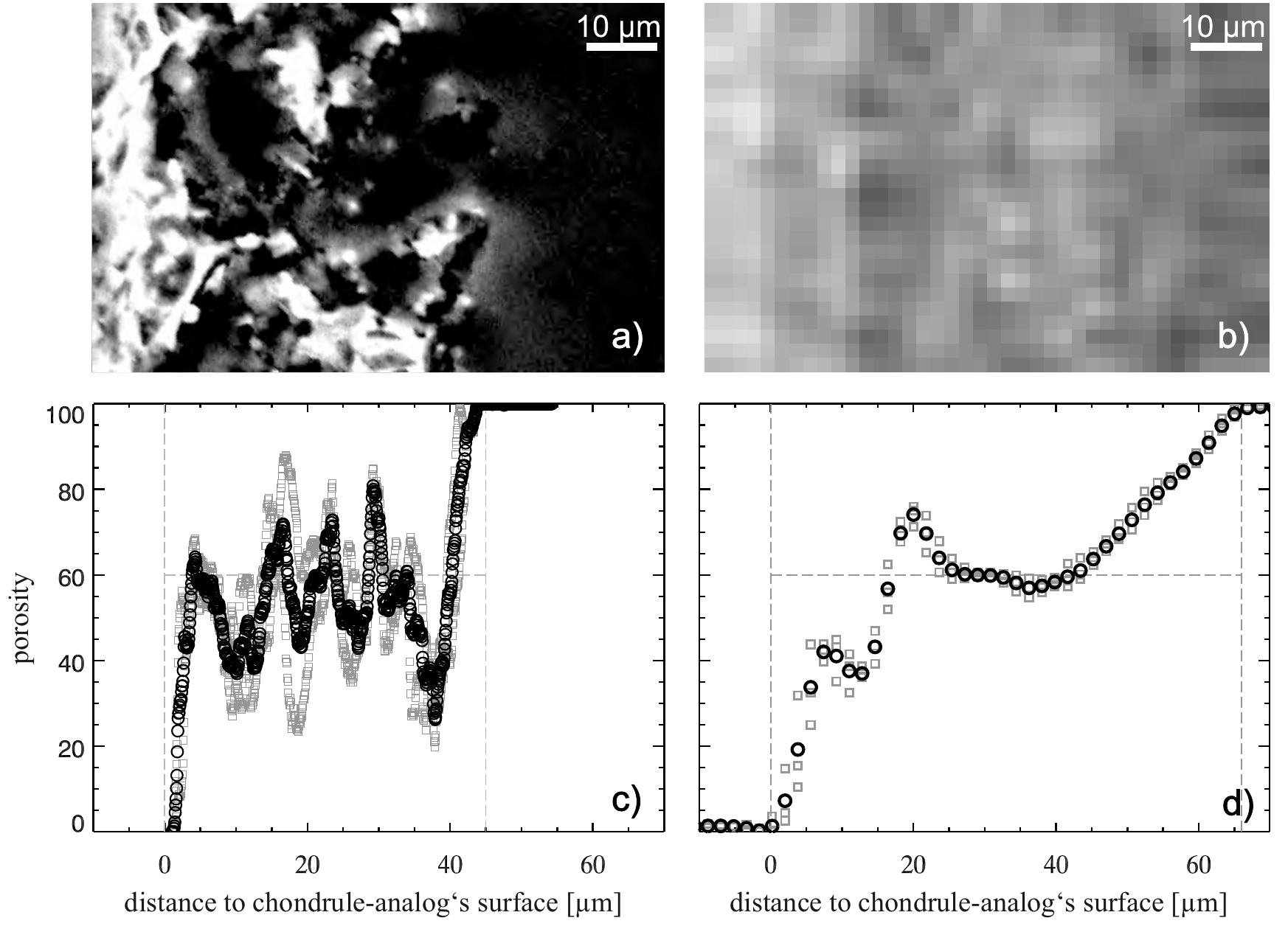}
                    \caption{\label{fig:fo3}The figure shows an analysis of the dust rim accreted to the Fo3-chondrule at elevated temperatures. This chondrule was first analyzed with the micro-CT at the Natural History Museum, London and was not embedded in epoxy resin for this measurement. Later it was embedded in epoxy resin, but it had lost a large part of its rim due to transportation. The embedded sample was then analyzed with an SEM. a) The BSE-image shows a small part of the Fo3-chondrule rim that was not completely destroyed and could be used for the measurement. The image shown in b) is a typical reconstructed micro-CT slide. The resolution of this image is lower by a factor of 29 compared with the BSE-image shown in a). A larger part of the micro-CT reconstruction of the Fo3-chondrule can be found as online material (electronic annex). The plots c) and d) show the results of the SEM and the micro-CT measurements of the dust rim porosity, respectively. Plotted is the rim porosity as a function of radial distance from the surfaces of the Fo3-chondrule analog. The porosity of the chondrule analog was normalized to a value of zero. A porosity value of 100 vol.\% denotes the outer border of the rim. Both, the outer and inner border of the rim are indicated by vertical dashed lines. The horizontal line at a porosity of 60 vol.\% depicts the mean value for the inner part of the accreted dust rim. The mean values of all measurements represented by the black circles, squares indicate single measurements.}
                \end{center}
            \end{figure*}

            \begin{figure*}[p]
                \begin{center}
                    \includegraphics[width=\textwidth]{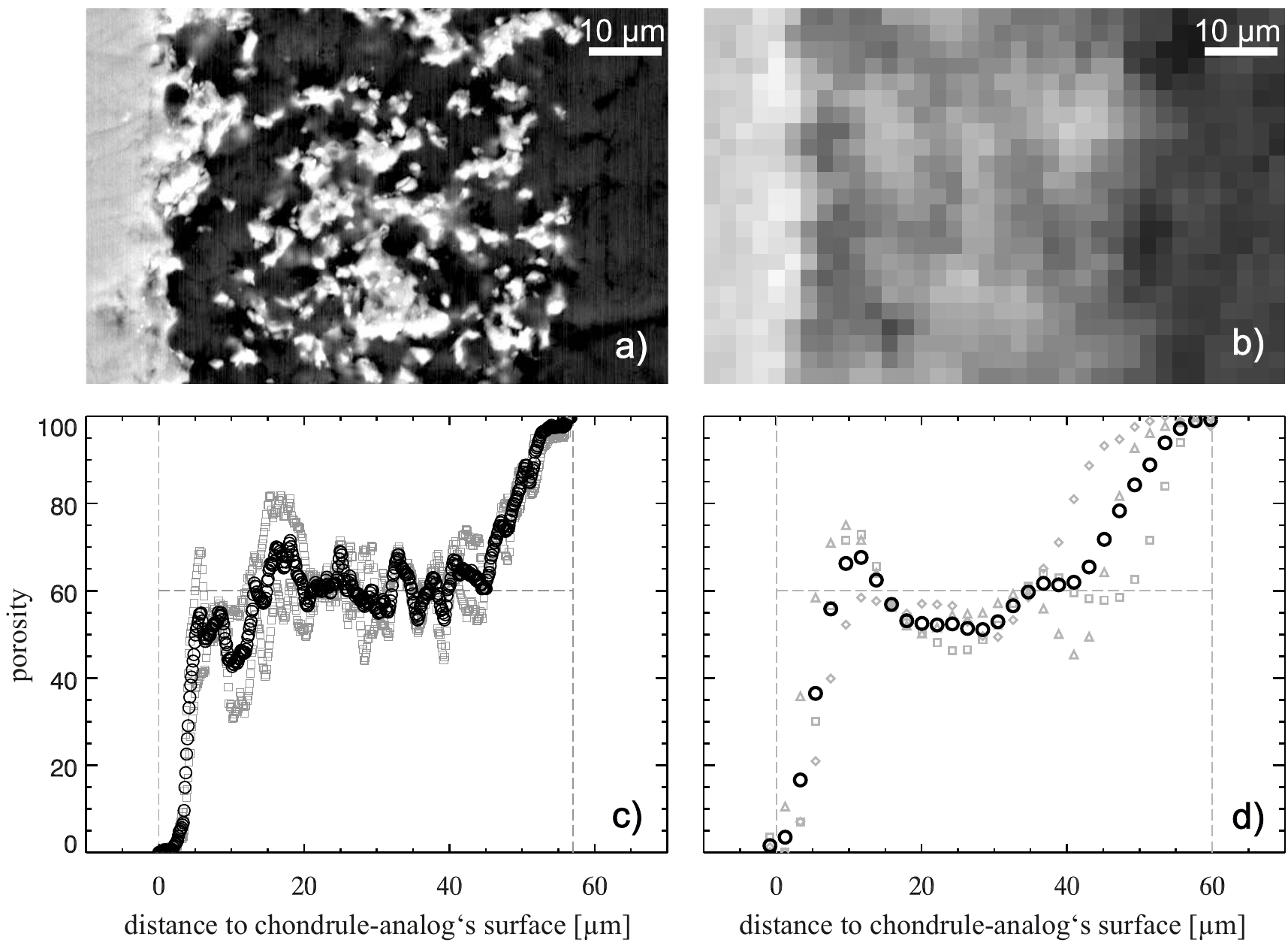}
                    \caption{\label{fig:sp1} Same as Figure \ref{fig:fo3} but for the Sp1-chondrule analog. The figure shows the analysis of the dust rim accreted to the Sp1-chondrule at elevated temperatures. This chondrule was first analyzed with the SEM at the TU Braunschweig and was therefore embedded in epoxy resin and polished to its midplane. Later it was carefully cut out of the epoxy resin block to be analyzed with a micro-CT at the TU Braunschweig. a) The BSE-image shows a typical part of the Sp1-chondrule rim. The image shown in b) is a typical reconstructed micro-CT slide. The resolution of this image is lower by a factor of 19 compared with the BSE-image shown in a). The plots c) and d) show the results of the SEM and the micro-CT measurement of the dust rim porosity, respectively. Plotted are the rim porosities as a function of radial distance from the surfaces of the Sp1-chondrule analog. The porosity of the chondrule analog was normalized to a value of zero. A porosity value of 100 vol.\% denotes the outer border of the rim. Both, the outer and inner border of the rim are indicated by vertical dashed lines. The horizontal line at a porosity of 60 vol.\% depicts the mean value for the inner part of the accreted dust rim. The mean values of all measurements are represented by black circles, squares indicate single measurements.}
                \end{center}
            \end{figure*}

\subsection*{Micro Computer Aided Tomography (micro-CT)}
The dust-coated Fo3-chondrule analog was scanned with a cabinet-based Nikon Metrology HMX ST 225 micro-CT scanner. The X-ray source produces a polychromatic beam and was operated in the range of 120-160 kV acceleration voltage and 60-100 $\mu$A current. X-ray transmission was recorded on a $2000 \times 2000$ pixel Perkin Elmer XRD 1621 AN3 HS detector panel with a 16 bit intensity depth. Reconstruction of the radial projection images into a stack of serial slices was carried out using the cone-beam, back-projection algorithms in the CT-Pro software by Nikon Metrology. Beam hardening was corrected for using an algorithm built in the reconstruction software. More details on measurement conditions can be found in \citet{Hezel2011GCAsub}. The hot Sp1-chondrule analog was scanned by a MicroXCT-400 (Manufacture: xradia), using an acceleration voltage of 80 kV and 125 $\mu$A current. A nanotom 160NF (Manufacture: phoenix) X-ray scanner was used to analyze Murchison. Measurement conditions were an acceleration voltage of 150 kV and a current of 28 $\mu$A.

\subsection*{Scanning Electron Microscopy (SEM)}
All three chondrule analogs were analyzed by scanning electron microscopy. They were embedded in epoxy resin and polished down to their midplane. The hot Sp1-chondrule analog was analyzed with the scanning electron microscope JSM 6400 (Manufacture: JEOL), the Fo3-chondrule analog as well as the cold  Sp1-chondrule analog were analyzed using a Cambridge S360 SEM. The details of the SEM measurements are summarized in Table \ref{table_method}. Both SEMs were operated with an acceleration voltage of 20 kV. Back-scattered electron (BSE) images were taken, for which the gray value represents the combined atomic weight for each pixel of the analyzed sample. Thus, the gray values of a BSE image is a measure of the density of the sample at each pixel. An additional advantage of this method is that the low-density epoxy resin appears dark on the BSE images, because of its low atomic weight.

\section{Results}

\subsection*{Porosity of dust rims on chondrule analogs}
The Fo3- and the hot Sp1-chondrule analogs were analyzed by micro-CT with a voxel edge length of 1.8 $\mathrm{\mu m}$ and 2.1 $\mathrm{\mu m}$, respectively. Thus, the resolution was in both cases too low to resolve all of the individual olivine dust grains so that the data analysis was done by averaging the gray values of the voxels as a function of distance to the chondrule analog's surface. To determine the rim porosity, three volumes of the dust rim, each having a size of $55 \times 65 \times 100$ voxels for the Fo3-chondrule analog and $30 \times 26 \times 45$ voxels for the Sp1-chondrule analog were arbitrarily chosen from the data set. These data cuboids were cut out of the chondrule analog in such a way that the face bending of the chondrule analog's surface was negligible. One scaled layer of such a data cuboid is shown in Figure \ref{fig:fo3}b) for the Fo3-chondrule analog and in Figure \ref{fig:sp1}b) for the hot Sp1-chondrule analog. The images were normalized by setting an upper and a lower threshold. For the porosity, the lower limit, i.e. 0 vol. \% porosity, was defined as the gray value representing the mass density of the olivine dust (3.2 $\mathrm{g\,cm^{-3}}$). This lower limit is present in both samples at the boundary between dust rim and chondrule analog where a dense layer of olivine dust was sintered to the chondrule analog's surface. The upper limit, i.e. 100 vol. \% porosity, is defined by the gray value representing the mass density of the background, i.e. air for the Fo3-chondrule analog and the epoxide resin for the hot Sp1-chondrule analog. To derive the porosity at any distance from the chondrule analog's surface, we averaged the normalized gray values for every line and over the complete stack of reconstructed images. The result for the three data cuboids for each chondrule is plotted in Figs. \ref{fig:fo3}d) and \ref{fig:sp1}d), respectively. The gray squares and triangles denote single measurements, the circles the average rim porosity as a function of the distance to the chondrule analog's surface. A horizontal dashed line indicates the mean porosity of the rim's center where we found a plateau at $\sim 60$ vol.\% porosity for both samples.

\begin{figure}[t]
    \begin{center}
        \includegraphics[width=9cm]{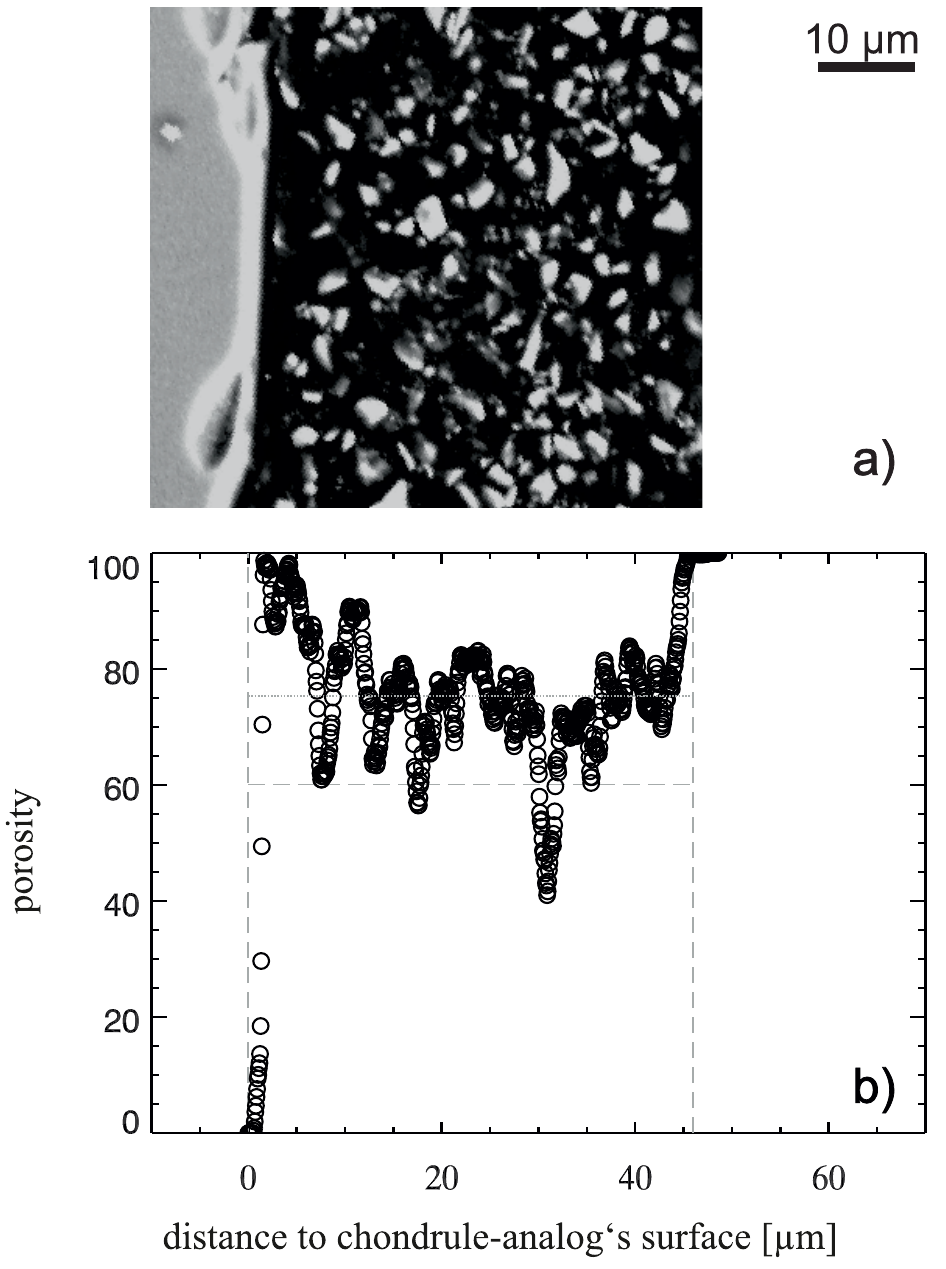}
        \caption{\label{fig:sp1-cold}Dust rim accreted to the Sp1-chondrule at room temperature. This chondrule was analyzed with the SEM at the TU Braunschweig. a) The BSE-image shows a part of the cold-accreted chondrule rim. b) shows the result of the rim porosity measurement. Plotted is the rim porosity as a function of the radial distance from the surface of the chondrule analog. The porosity of the chondrule analog was normalized to a value of zero. A porosity value of 100 vol.\% denotes the outer border of the rim. The outer and inner border of the rim are indicated by vertical dashed lines. The dashed horizontal line at a porosity of 60 vol.\% depicts the mean value that was found for the hot coated chondrules shown in Figure \ref{fig:fo3} and Figure \ref{fig:sp1}, the dotted line depicts the mean porosity of 75 vol.\% for the cold accreted rim. }
    \end{center}
\end{figure}

We cross-checked the porosity results from micro-CT to porosity results determined from BSE-images. This was done in the same way as above, only that BSE-images are used instead of a tomographic slice. Two BSE images were used to determine porosities for the hot Sp1-chondrule analog and one image each was used to determine porosities the Fo3- and the cold Sp1-chondrule analog. The results for all three chondrule analogs are plotted in Figure \ref{fig:sp1}c), Figure \ref{fig:fo3}c) and Figure \ref{fig:sp1-cold}b), respectively. Most of the Fo3-chondrule rim was destroyed during transportation and therefore only the inner part of the rim could be analyzed. An image of the remaining rim is shown in Figure \ref{fig:fo3}a). This image has a resolution of 0.06 $\mathrm{\mu m}$ per pixel. The rim accreted by the hot Sp1-chondrule analog is shown in the BSE image of Fig. \ref{fig:sp1}a). The image has a resolution of 0.11 $\mathrm{\mu m}$ per pixel. Thus, most of the olivine dust particles could be resolved in detail.

Fig. \ref{fig:sp1-cold}a) shows the dust rim accreted by the cold-coated chondrule. The image has a resolution of 0.06 $\mathrm{\mu m}$ per pixel.

The dust rim porosities for the Fo3- and hot Sp1-chondrule analog are plotted as a function of the distance to the chondrule analogs' surface in Figure \ref{fig:fo3} and Figure \ref{fig:sp1}. We found that the results of the two analysis techniques we used are in good agreement, giving identical values for the mean porosity of the dust rims of $\sim 60$ vol. \%. The results from micro-CT display a smooth porosity curve, due to the comparatively low spatial resolution and the large number of images we used for the analysis. The porosity curves resulting from the BSE-images show considerably more scattering because of the intrinsically high spatial resolution at which contributions from individual dust grains become important, and the low number of images used for the SEM measurements. Nevertheless, all curves possess the same overall shape and yield the same porosity value of 60 vol. \% for the main part of the rim.

A general difference of the accretion efficiency among the two chondrule analogs were not found in the experiments. For the Fo3-chondrule analog we measure a rim thickness of $\sim \mathrm {66 \,\mu m}$ using micro-CT compared to a much lower value using SEM. This is because the chondrule lost most of its rim during transportation. For the hot Sp1-chondrule we found a rim thickness of $\sim \mathrm {60\, \mu m}$ using the micro-CT technique and $\sim \mathrm {57\,\mu m}$ with SEM. The rim thicknesses normalized to the core diameters are much lower in this study ($\sim 0.04$) than those measured for fine grained rims by \citet{MetzlerEtal:1992} ($\sim 0.19$). This difference depends on the one hand on the coating technique that limits the thickness of the rim due to the levitation height in the funnel, and on the other hand on the limited coating time in the experiment. \citet{BeitzEtal:2011b} have shown that the rim thickness depends linearly on the coating time and could be identified with the time a chondrule is flying freely in the nebula. To constrain whether the gas-to-dust ratio is comparable to that expected for the solar nebula, we measured the dust flux, as well as the gas flux through the funnel, and calculated the gas-to-dust ratio to be between 400 and 600, which is a comparable to the value used in simulations by \citet{morris&desch:2010} within a factor of two.

\begin{figure}[t]
    \begin{center}
        \includegraphics[width=9cm]{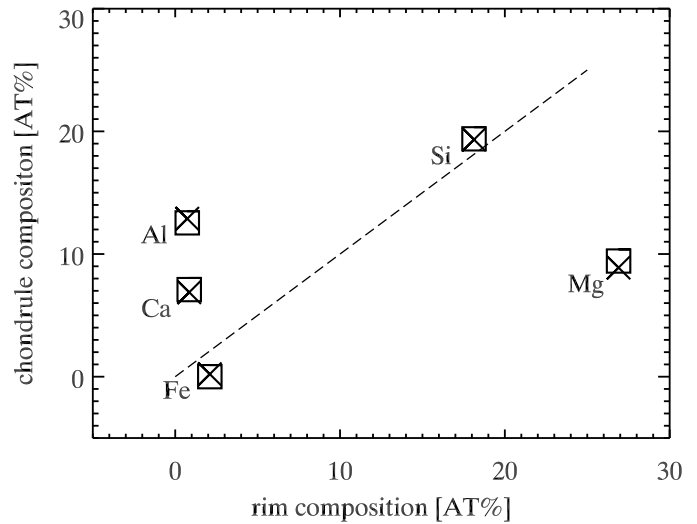}
        \caption{\label{fig:compo} Elemental composition at the chondrule analog's center (squares) and close to its surface (crosses) compared to the rim composition measured at the outer border of the rim. The dashed lines can be interpreted as the direction in which the crosses should have migrated if elemental exchange between the two components occurred in a possible reaction zone between the rim (San Carlos olivine composition) and the chondrule analog material. }
    \end{center}
\end{figure}

The BSE images show whether or not the chondrule analogs were crystalline after the experimental runs. It has not been the focus of this study to reproduce realistic chondrule textures. Thus, realistic textures could not be expected regarding the experimental temperature profile we used. We found that the hot Sp1-chondrule analog is glass. This allows us to analyze the Sp1-chondrule analog's composition using EDX and to test whether or not elemental exchange between the chondrule analog and its rim occurred.

The BSE-images of Figure \ref{fig:fo3}a) and \ref{fig:sp1}a) show that several dust grains are sintered to the surface of the chondrule. The experimental temperature of $1100^\circ$C was well below the melting point of both the chondrule analogs and the olivine dust ($\mathrm{T_{liq}\geq 1750^\circ}$C). To constrain whether this sintering is accompanied by an elemental exchange between dust grains and chondrule analog, we measured the element composition of the different components at this interface using the SEM. The measurement was done on the hot Sp1-chondrule, which was still a glass. Three measurements were performed, one on the chondrule analog side of the chondrule-rim boundary (a few $\rm \mu$m inside of the chondrule analog's edge), the second at the center of the chondrule analog to measure the unaltered composition of the chondrule analog, and finally, the natural composition of the olivine dust was determined by a spot analysis of a large olivine grain well outside the chondrule analog-rim boundary. In Figure \ref{fig:compo}, the measured compositions of the chondrule analog is plotted against the olivine dust composition. The squares denote the composition of the chondrule analog at its center, whereas the crosses show the composition close to the chondrule analog-dust boundary. A comparison of the data from the chondrule analog's center and its periphery shows no sign of elemental exchange with material from the olivine dust. A potential diffusion can be best analyzed using the constituents Fe and Mg as tracers, because both diffuse easily. The chondrule analog contains no Fe and has a low Mg content. Both elements are much higher concentrated in the olivine dust rim material so that a diffusion of these elements should lead to an enhancement in their abundances close to the chondrule analog's surface. However, there is no difference in concentrations of Fe and Mg between the chondrule analog in the center and at the border of the rim. Therefore, we can exclude a substantial exchange of these elements between the two components.

\subsection*{Porosity of chondrule rims in Murchison}
\begin{figure}[t]
    \begin{center}
        \includegraphics[width=9cm]{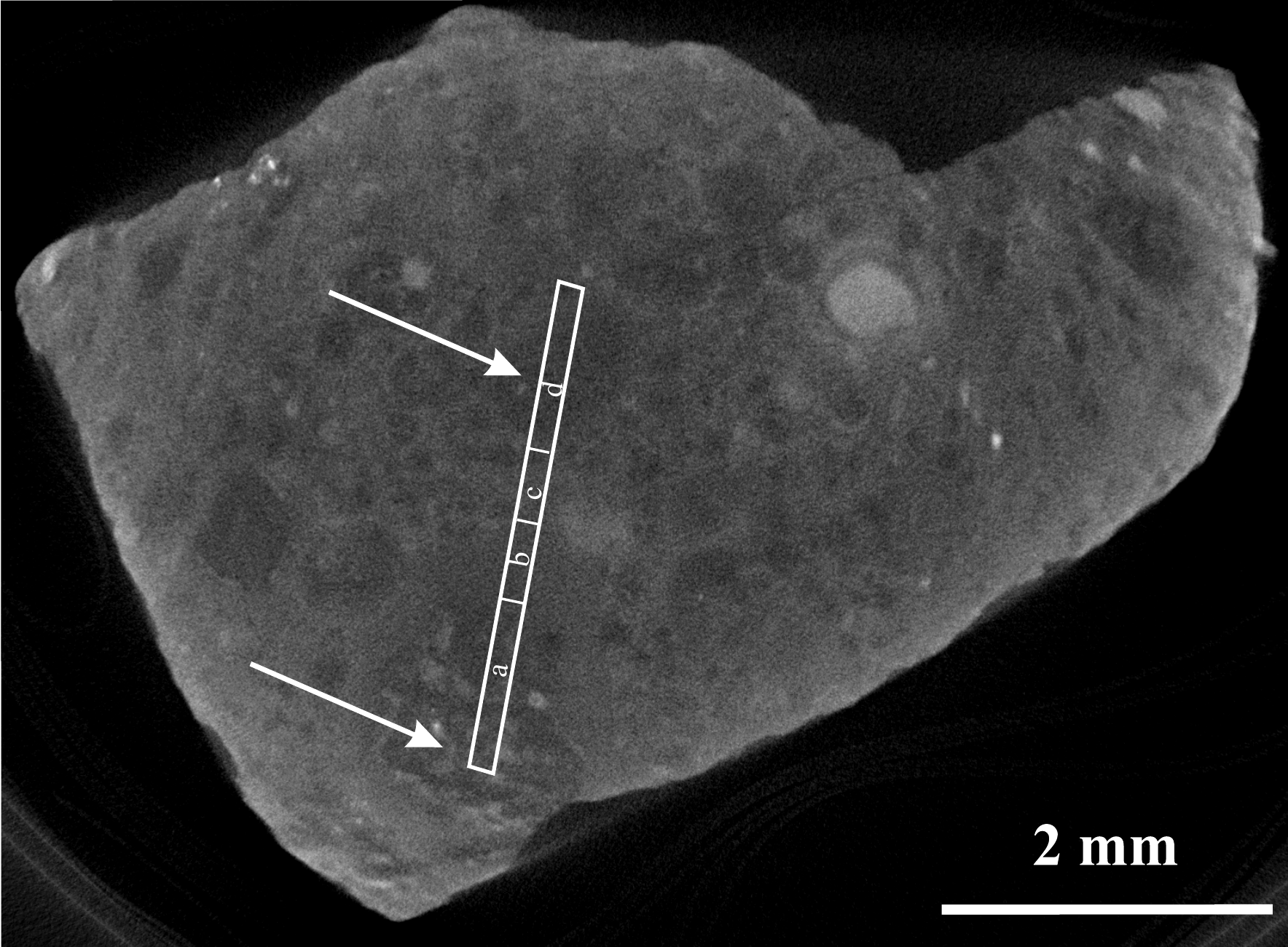}
        \caption{\label{fig:image_Murchison} The image shows a typical CT-slice of the analyzed piece of Murchison. The white box highlights the area used to analyze the porosity of the chondrules indicated with (a) and (d) as well as the porosity of the rims indicated by (b) and (c), respectively. The resolution of the image is 5 $\mathrm{\frac{\mu m}{voxel}}$, the white arrows indicate chondrules.  }
    \end{center}
\end{figure}

\begin{figure}[t]
    \begin{center}
        \includegraphics[width=9cm]{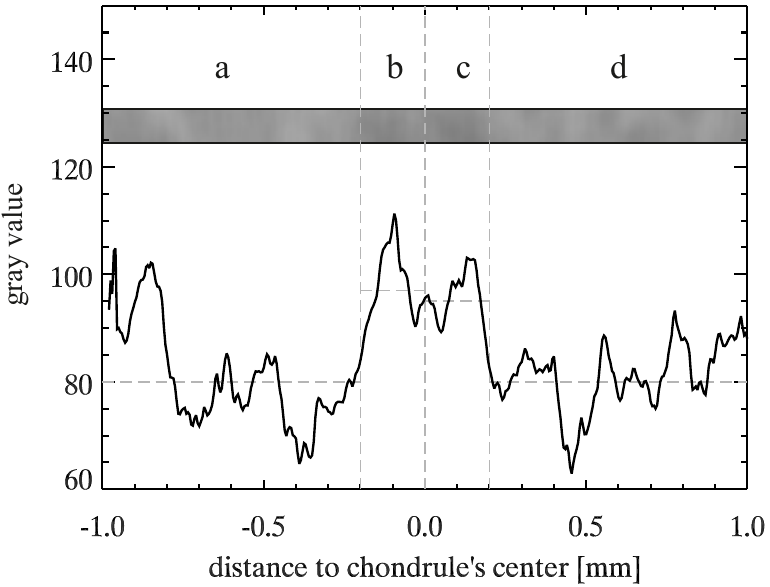}
        \caption{\label{fig:Plot_Murchison} The figure shows the gray value distribution of a piece of Murchison meteorite. The inset shows a stack of the 20 micro-CT slices used for the measurement and its alignment in the meteorite is represented by the white box in Figure \ref{fig:image_Murchison}. The chondrules are indicated by the brighter areas (a) and (d), and their mean gray value refers to a porosity of zero. The areas labeled with (b) and (c) show the fine-grained chondrule rim of the two chondrules (a) and (d), respectively. The difference in gray values between the mean chondrule gray value and the rims gray values possess values of 17 and 15, respectively. The vertical gray lines depict the transitions between the different regions, whereas the horizontal line denotes the mean gray values. The porosity of the rims is calculated from these differences to be $\sim$ 10 vol. \%}
    \end{center}
\end{figure}

To compare the rim porosities from the experimental samples with porosities of real chondrule rims, we scanned a piece of Murchison using the micro-CT at the Deutsches Zentrum f\"ur Luft- und Raumfahrt (DLR) in K\"oln. This allows us to compare our results to the porosity measurements of CM chondrites by \citet{Ashworth1977}. The volume to measure the porosity of Murchison chondrule rims had a dimension of  $472 \times 22 \times 20$ voxels with a voxel edge length of 5 $\mathrm{\mu m}$. The alignment of the measurement volume was chosen such that two chondrules are at the sides of the volume and their dust rims are in the center (see white box in Figure \ref{fig:image_Murchison}). The normalization of the images was done in the same way as described above for the rimmed chondrule analogs. The distribution of the gray values for this data cuboid are shown in Fig. \ref{fig:Plot_Murchison}. These gray values are interpreted such that they represent the porosity distribution in this cuboid. Therefore, a constant density for the chondrules and the rims have to be assumed. Thus, we interpret the chondrules as igneous beads with no porosity and these are normalized to a porosity of zero. The upper threshold referring to a porosity of 100 vol. \% is given by the background noise outside of the chondrite. Using these assumptions we calculated a mean porosity of the chondrule rims of $\rm\sim 10 \,vol. \%$.

\section{Discussion}

\begin{table*}[t]
\center
    \caption{Degree of compaction of four CV3 chondrites relative to an uncompacted matrix porosity of 60 vol.\%. }
    \label{table_compaction}
\begin{tabular}{c c c c c }
\hline
     CV3 Chondrite & matrix abundance & porosity  & difference between measured  &Compaction coefficient  \\
    &  [$\rm vol. \%$] & [$\rm vol. \%$]& and calculated value [$\rm vol. \%$]&  $\rm \zeta_{60 \,vol. \%}$  (see Eq. \ref{cczeta})\\
\hline

 Allende & 63.1$^\flat$ & 23$^\natural$& 14.9 & 0.39 \\
   Mokoia & 55.6$^\flat$ & 24$^\sharp$  &9.3 & 0.28\\
  Vigarano & 33.4$^\dag$ & 13$^\natural$ & 14.0& 0.70\\
  Leoville & 31.7$^\dag$ & 6$^\natural$ & 6.0& 0.31\\
  \hline
\end{tabular}\\
\footnotesize
$^\flat$ fraction matrix and porosity \citep{Hezel2011GCAsub}, $^\dag$ fraction matrix \citep{ebel:2009},$^\natural$ chondrite porosities taken from \citet{Consolmagno2008}, $^\sharp$ chondrite porosities taken from \citet{corriganEtal:1997}
\end{table*}

We used two independent techniques to study experimentally produced dust rims and proved that micro-CT delivers sufficient resolution to study the porosity of such rims. The porosity of the dust rims around the chondrule analogs are very low at the boundary between chondrule analog and dust rim, due to partial melting or sintering. The porosity rapidly increases to a mean value of 60 vol.\% at about half way to the outer boundary of the dust rim. At the outer boundary, the porosity increases from the mean value of 60 vol.\% to a porosity of 100 vol.\% within only of 10 to 20 $\rm \mu m$.

The dust rim itself is not a layer of homogeneous thickness, and varies by several \mum. This variation could be the result of rotation of the chondrule analog while aerodynamically levitated in the funnel. Another reason could be shearing-off of the outer rim parts during the embedding process into the epoxy resin, but the variation can also be explained by the stochastic absence/presence of a few larger olivine grains. There is no apparent evidence that the epoxy resin changed the porosity dust rims. For both hot coated chondrule analogs, a consistent depression of rim porosity could be found close to the inner boundary of the dust rim between 5 and 20 $\mu$m from the chondrule analog's surface. This gap in dust packing, easily visible in the images and the porosity curves in Figures \ref{fig:fo3} and \ref{fig:sp1}, can be explained by a shrinkage of the chondrule analogs during their rapid cooling. This is caused by abruptly switching off the laser beam, and seems to be an artefact of the experimental run and not a specific physical process acting during dust rim formation.

Most part of the dust rims produced at 1100 $^\circ$C possesses a constant porosity of 60 vol.\% and, hence, can be regarded as the typical result of the low-velocity accretion process of polydisperse dust at these temperatures.
\citet{Consolmagno2008} reviewed the porosities of ordinary and carbonaceous chondrites and found a wide range of bulk chondrite porosities between 9.7 and 35 vol.\% for carbonaceous chondrites and between 7.0 and 8.2 vol.\% for ordinary chondrites (cf. their Tables 1 and 4; see also \citet{FlynnEtal:1999}). These numbers represent the macro-porosity for the overall meteorites.

The cold Sp1-chondrule analog has a porosity of 75 vol.\% at the center of the dust rim, which is higher than the measured porosity of the hot coated samples. It appears that at higher temperature rims of lower porosities form by sintering effects of the dust grains. Chondrites of the CV3 type have rims with low porosities and might therefore have formed at higher temperatures than the dust rims around our chondrule analogs. Nevertheless, the fine grained rims in CM meteorites are typically not altered due to temperature effects or sintering of dust grains and therefore, the dust rim accreted by the cold-chondrule should be compared with the fine grained rims found in CM chondrites \citep{MetzlerEtal:1992}.

We determined a porosity of $\sim 10$ vol.\% for the dust rim around chondrules in Murchsion, which is within the range reported by \citet{Ashworth1977} (6 to 15 vol.\%). The porosity value is calculated by the difference in gray values of the micro-CT images, assuming the same density of chondrule and rim material. The constant density of rim and chondrule material must be seen as a rough estimation and the so calculated porosity is influenced by variations in the density of the dust grains and the chondrules. These grain-density variations are measured by \citet{Macke2011} to be about 6 vol.\% for 14 pieces of Murchison, using helium ideal-gas pycnometry. But the grain can vary locally even more due to enclosed sulfides or Fe rich grains. Taking the errors from \citet{Macke2011}, our results are still within the range of rim porosities given by \citet{Ashworth1977}. Therefore, we conclude that the method to derive the density characteristics by means of the gray values in micro-CT analysis assuming a constant grain density is a well suited method to estimate the global porosity structure and is consistent with the mean density of other methods.

There is a significant difference in porosities of dust rims around the chondrule analogs (60 and 75 vol.\%, respectively) and the Murchison chondrule rims ($\sim 10$ vol.\%). If the rims around the Murchison chondrules formed in a similar way to our experimental dust rims, significant secondary compaction and/or alteration is required to reduce the initial dust rim. There are several possibilities to achieve this: i) As shown above, higher temperatures during dust rim formation produces rims of lower porosities, but the rim will possibly not possess a fine grained structure any more, because of sintering of the dust grains. ii) Gentle compaction due to impacts can decrease the porosity \citep{wasson1995}, but this is limited to the random close packing and can also not explain the reported porosities below this packing limit. iii) Aqueous alteration of phyllosilicates is coupled to an increase of the dust grain volume and could lead to a decrease in porosity of $\sim 10$ vol.\% \citep{WilsonEtal1999}.

In this work, we have shown that accretion of dust rims in low-velocity collisions leads to a high porosity, which is not found in chondrites so that we conclude that additional compaction and alteration is required. Possible processes need to be investigated in further experiments.

\citet{BeitzEtal:2011b} showed that dust covered mm-sized glass beads stick to one another in low-velocity collisions and are likely to form clusters. Assuming dust rimmed chondrules act as these glass beads and clusters, once formed, chondrule clusters could further accrete nebular dust grains and the empty space between the dust rimmed chondrules could further be filled with dust and with a similar porosity as the dust rims. Finally, this very basic scenario, in which no compaction in collisions occurs, produces a big chondrite-like parent body.  We calculate the bulk porosity of such a parent body assuming that matrix has the porosity of the dust rims around our chondrule analogs and all non-matrix components have no significant porosity. We compare the model porosity of such a parent body to four CV chondrites (Allende, Mokoia, Vigarano and Leoville), as there is a large scatter of reported porosities and component modal abundances, even in a single subgroup of chondrites.

\begin{figure}[t]
    \begin{center}
        \includegraphics[width=9cm]{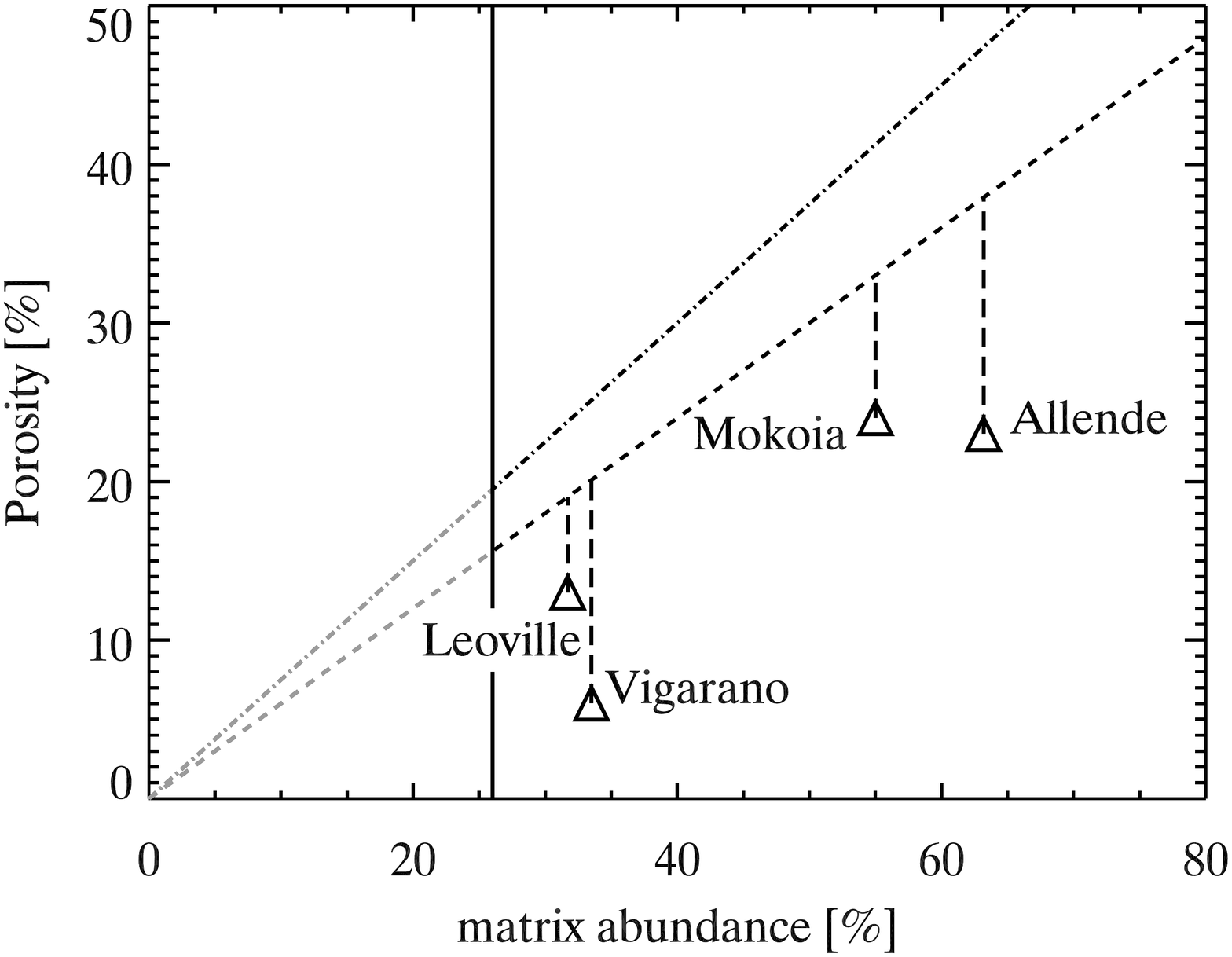}
        \caption{\label{fig:compaction} The measured porosities for four different CV3 chondrites against their matrix abundances (data from \citet{ebel:2009} and \citet{Hezel2011GCAsub}). The values of the measured porosities of 23 vol.\% for Allende, 13 vol.\% for Leoville and 6 vol.\% for Vigarano are taken from \citet{Consolmagno2008} and the porosity of 24 vol.\% for Mokoia is taken from \citet{corriganEtal:1997}. The dashed and the dashed-dotted lines denote the bulk porosity of a hypothetical chondrite with a matrix porosity of 60 vol.\% and 75 vol.\%, respectively. The difference between the measured porosities and the dashed line for each chondrite is a measure for the degree of compaction that the parent body experienced. The model is not valid on the left hand side of the vertical line, which denotes random close packing for identical spherical particles, because a higher chondrule abundance is not achievable without a deformation or breaking of the chondrules and such an energetic process is excluded in the hypothesis.}
    \end{center}
\end{figure}

First, bulk porosities of the chondrites were calculated using the results from our experiments, later they will be compared to the bulk porosities found in the literature for these chondrites. For this calculation, we assumed that the non-matrix fraction of the chondrites has no significant porosity (although there is a low porosity in such components, cf. \citet{Hezel2011GCAsub}) and that the porosity is restricted to the matrix. This means that the bulk porosity of a meteorite is only dependent on the volume fraction of the matrix. For the matrix porosity, we assume the porosities measured for the hot and cold chondrule rims to be a measure for the porosity of a uncompacted matrix fraction of a chondrite. Thus, the dashed and dash dotted lines in Figure \ref{fig:compaction} denote the expected porosity of a hypothetical chondrite with matrix porosities of 60 vol.\% and 75 vol.\%, respectively. Also shown are the component modal abundances taken from \citet{Hezel2011GCAsub} for Allende and Mokoia and from \citet{ebel:2009} for Vigarano and Leoville. These measured porosities (triangles) are lower than the calculated in all four chondrites, which we interpret as being due to compaction processes on their parent bodies or further low temperature alteration. The length of the dashed line between the calculated and the measured porosities is a measure of the degree of compaction. However, there is no correlation between the degree of compaction and the porosity, whereas the matrix abundance correlates positively with increasing bulk chondrite porosity.

A compaction coefficient, defined as
\begin{equation}\label{cczeta}
    \zeta_{60\,\mathrm{vol.} \%}= 1-\frac{P_{meas}}{P_{calc,60\, \mathrm{vol.} \%}} ,
\end{equation}
with $P_{meas}$ and $P_{calc,60\,\mathrm{vol.} \%}$ being the measured and modeled porosities (in this case, the latter is assumed to be 60 vol.\%), can be used as a measure of the degree of compaction of the chondrites. That numerical values of the compaction coefficient fall between 0 and 1 and can be interpreted as the degree of compaction that does not depend on the matrix abundance anymore. We use the compaction coefficient to compare the degree of compaction of different chondrites relative to a matrix porosity of 60 vol.\%. All values for the conducted chondrites are summarized in Table \ref{table_compaction}.

The mean compaction is 10.8 vol.\% for the four chondrites. However, the porosity difference for each chondrite scatters between only 6 vol.\% for Leoville and 14.9 vol.\% for Allende. A comparison of the compaction coefficients of the four chondrites shows that Mokoia ($\rm \zeta_{60\,vol.\%}=0.28$) is the least compacted, Leoville ($\rm \zeta_{60\,vol.\%}=0.31$) and Allende ($\rm \zeta_{60\,vol.\%}=0.39$) are intermediate compacted and Vigarano ($\rm \zeta_{60\,vol.\%}=0.70$) is the most altered or compacted chondrite in our study. Nevertheless, a significant compaction or alteration must have happened to all tested CV chondrites.

\subsection*{Acknowledgements}
E.B. thanks the Deutsche Forschungsgemeinschaft (DFG) for support under grant Bl 298/13-1 as part of the SPP 1385 "The first 10 Millon Years of the Solar System". We thank Birgit Gerke from the Institut f\"ur theoretische und physikalische Chemie (TU Braunschweig) and Johannes Ledig form the Institut f\"ur Halbleitertechnik (TU Braunschweig) for providing us with the BSE-images and composition measurements. We also thank Stephan Olliges from the Institut f\"ur Partikeltechnik (TU Braunschweig) and Stefan Frank-Richter and Matthias Sperl from the Deutsches Zentrum für Luft- und Raumfahrt (DLR) in K\"oln for providing us with CT measurements. We are grateful to our referee Roger H. Hewins, the two anonymous reviewers, and our AE Sara Russell for their valuable input and critiques.

\bibliographystyle{apalike}
\bibliography{literatur}

\begin{thebibliography}{}

\bibitem[Alexander et~al., 2008]{AlexanderEtal:2008}
Alexander, C. M.~O., Grossman, J.~N., Ebel, D.~S., and Ciesla, F.~J. (2008).
\newblock The formation conditions of chondrules and chondrites.
\newblock {\em Science}, 320(5883):1617--1619.

\bibitem[Allen et~al., 1980]{Allen1980}
Allen, J.~S., Nozette, S., and Wilkening, L.~L. (1980).
\newblock A study of chondrule rims and chondrule irradiation records in
  unequilibrated ordinary chondrites.
\newblock {\em Geochimica et Cosmochimica Acta}, 44(8):1161 -- 1175.

\bibitem[{Amelin} et~al., 2002]{AmelinEtal:2002}
{Amelin}, Y., {Krot}, A.~N., {Hutcheon}, I.~D., and {Ulyanov}, A.~A. (2002).
\newblock {Lead isotopic ages of chondrules and Calcium-Aluminum-rich
  inclusions}.
\newblock {\em Science}, 297:1678--1683.

\bibitem[Ashworth, 1977]{Ashworth1977}
Ashworth, J. (1977).
\newblock Matrix textures in unequilibrated ordinary chondrites.
\newblock {\em Earth and Planetary Science Letters}, 35(1):25 -- 34.

\bibitem[Beitz et~al., 2012]{BeitzEtal:2011b}
Beitz, E., Güttler, C., Weidling, R., and Blum, J. (2012).
\newblock Free collisions in a microgravity many-particle experiment – ii: The
  collision dynamics of dust-coated chondrules.
\newblock {\em Icarus}, 218(1):701 -- 706.

\bibitem[{Bischoff} and {Geiger}, 1995]{bischoff_geiger:1995}
{Bischoff}, A. and {Geiger}, T. (1995).
\newblock {Meteorites for the Sahara: Find locations, shock classification,
  degree of weathering and pairing}.
\newblock {\em Meteoritics}, 30:113--122.

\bibitem[{Blum} and {Schr{\"a}pler}, 2004]{BlumSchraepler:2004}
{Blum}, J. and {Schr{\"a}pler}, R. (2004).
\newblock {Structure and Mechanical Properties of High-Porosity Macroscopic
  Agglomerates Formed by Random Ballistic Deposition}.
\newblock {\em \prl}, 93(11):115503.

\bibitem[Brearley and Jones, 1998]{Brearley&Jones:1998}
Brearley, A. and Jones, R. (1998).
\newblock Chondritic meteorites.
\newblock {\em Reviews in Mineralogy and Geochemistry}, 36(1):3--1.

\bibitem[Carballido, 2011]{Carballido2011876}
Carballido, A. (2011).
\newblock Accretion of dust by chondrules in a mhd-turbulent solar nebula.
\newblock {\em Icarus}, 211(1):876 -- 884.

\bibitem[Consolmagno et~al., 2008]{Consolmagno2008}
Consolmagno, G., Britt, D., and Macke, R. (2008).
\newblock The significance of meteorite density and porosity.
\newblock {\em Chemie der Erde - Geochemistry}, 68(1):1 -- 29.

\bibitem[{Corrigan} et~al., 1997]{corriganEtal:1997}
{Corrigan}, C.~M., {Zolensky}, M.~E., {Dahl}, J., {Long}, M., {Weir}, J.,
  {Sapp}, C., and {Burkett}, P.~J. (1997).
\newblock {The porosity and permeability of chondritic meteorites and
  interplanetary dust particles}.
\newblock {\em Meteoritics and Planetary Science}, 32:509--515.

\bibitem[Cuzzi, 2004]{Cuzzi2004}
Cuzzi, J.~N. (2004).
\newblock Blowing in the wind: Iii. accretion of dust rims by chondrule-sized
  particles in a turbulent protoplanetary nebula.
\newblock {\em Icarus}, 168(2):484 -- 497.

\bibitem[{Desch} and {Connolly}, 2002]{DeschConnolly:2002}
{Desch}, S.~J. and {Connolly}, Jr., H.~C. (2002).
\newblock {A model of the thermal processing of particles in solar nebula
  shocks: Application to the cooling rates of chondrules}.
\newblock {\em Meteoritics and Planetary Science}, 37:183--207.

\bibitem[{Ebel} et~al., 2008]{ebel:2009}
{Ebel}, D.~S., {Brunner}, C.~E., and {Weisberg}, M.~K. (2008).
\newblock {Multiscale Abundance and Size Distribution of Inclusions in the
  Allende CV3 Meteorite by X-Ray Image Analysis of Slabs}.
\newblock In {\em Lunar and Planetary Institute Science Conference Abstracts},
  volume~39, pages 21--21.

\bibitem[Ebel and Grossman, 2000]{Ebel2000}
Ebel, D.~S. and Grossman, L. (2000).
\newblock Condensation in dust-enriched systems.
\newblock {\em Geochimica et Cosmochimica Acta}, 64(2):339 -- 366.

\bibitem[Flynn et~al., 1999]{FlynnEtal:1999}
Flynn, G.~J., Moore, L.~B., and Klöck, W. (1999).
\newblock Density and porosity of stone meteorites: Implications for the
  density, porosity, cratering, and collisional disruption of asteroids.
\newblock {\em Icarus}, 142(1):97 -- 105.

\bibitem[Gundlach et~al., 2011]{GundlachEtal:2011}
Gundlach, B., Kilias, S., Beitz, E., and Blum, J. (2011).
\newblock Micrometer-sized ice particles for planetary-science experiments - i.
  preparation, critical rolling friction force, and specific surface energy.
\newblock {\em Icarus}, 214(2):717 -- 723.

\bibitem[Hezel et~al., 2011]{Hezel2011GCAsub}
Hezel, D., Elangovan, P., Viehmann, S., Howard, L., Abel, R., and Armstrong, R.
  (accepted\setbox0=\hbox{2011}).
\newblock Visualisation and quantification of cv chondrites using
  micro-tomography.
\newblock {\em Geochimica et Cosmochimica Acta}.

\bibitem[{Hezel} et~al., 2008]{HezelEtal:2008}
{Hezel}, D.~C., {Russell}, S.~S., {Ross}, A.~J., and {Kearsley}, A.~T. (2008).
\newblock {Modal abundances of CAIs: Implications for bulk chondrite element
  abundances and fractionations}.
\newblock {\em Meteoritics and Planetary Science}, 43:1879--1894.

\bibitem[{Kring}, 1989]{Kring1989PhDT}
{Kring}, D.~A. (1989).
\newblock {\em {The Petrology of Meteoritic Chrondrules: Evidence for
  Fluctuating Conditions in the Solar Nebula.}}
\newblock PhD thesis, HARVARD UNIVERSITY.

\bibitem[Kring, 1991]{Kring1991}
Kring, D.~A. (1991).
\newblock High temperature rims around chondrules in primitive chondrites:
  evidence for fluctuating conditions in the solar nebula.
\newblock {\em Earth and Planetary Science Letters}, 105:65 -- 80.

\bibitem[Krot and Wasson, 1995]{Krot:1995}
Krot, A.~N. and Wasson, J.~T. (1995).
\newblock Igneous rims on low-feo and high-feo chondrules in ordinary
  chondrites.
\newblock {\em Geochimica et Cosmochimica Acta}, 59:4951 -- 4966.

\bibitem[Kurahashi et~al., 2008]{Kurahashi2008}
Kurahashi, E., Kita, N.~T., Nagahara, H., and Morishita, Y. (2008).
\newblock 26al–26mg systematics of chondrules in a primitive co chondrite.
\newblock {\em Geochimica et Cosmochimica Acta}, 72(15):3865 -- 3882.

\bibitem[{Lodders}, 2003]{Lodders2003}
{Lodders}, K. (2003).
\newblock {Solar System Abundances and Condensation Temperatures of the
  Elements}.
\newblock {\em \apj}, 591:1220--1247.

\bibitem[{Macke} et~al., 2011]{Macke2011}
{Macke}, R.~J., {Consolmagno}, G.~J., and {Britt}, D.~T. (2011).
\newblock {Density, porosity, and magnetic susceptibility of carbonaceous
  chondrites}.
\newblock {\em Meteoritics and Planetary Science}, 46:1842--1862.

\bibitem[Mathieu, 2009]{mathieu_phd}
Mathieu, R. (2009).
\newblock {\em Sodium solubility in silicate melts}.
\newblock PhD thesis, Institut National Polytechnique de Lorraine, Nancy,
  France.

\bibitem[Mathieu et~al., 2011]{MathieuEtal:2011}
Mathieu, R., Libourel, G., Deloule, E., Tissandier, L., Rapin, C., and Podor,
  R. (2011).
\newblock {$\rm Na_2O$ solubility in $\rm CaO–MgO–SiO_2$ melts}.
\newblock {\em Geochimica et Cosmochimica Acta}, 75(2):608 -- 628.

\bibitem[Mathieu and Pack, 2011]{mathieuandpack:2011}
Mathieu, R. and Pack, A. (2011).
\newblock Constraining chondrule formation using an aerodynamic levitation
  apparatus.
\newblock In {\em Lunar and Planetary Institute Science Conference Abstracts},
  volume~42, pages 2476--2476.

\bibitem[{McSween}, 1977]{McSween:1977}
{McSween}, jr, H.~Y. (1977).
\newblock {Carbonaceous chondrites of the Ornans type: A metamorphic sequence}.
\newblock {\em \gca}, 41:447--491.

\bibitem[{Metzler} et~al., 1992]{MetzlerEtal:1992}
{Metzler}, K., {Bischoff}, A., and {Stoeffler}, D. (1992).
\newblock {Accretionary dust mantles in CM chondrites - Evidence for solar
  nebula processes}.
\newblock {\em \gca}, 56:2873--2897.

\bibitem[{Morris} and {Desch}, 2010]{morris&desch:2010}
{Morris}, M.~A. and {Desch}, S.~J. (2010).
\newblock {Thermal Histories of Chondrules in Solar Nebula Shocks}.
\newblock {\em \apj}, 722:1474--1494.

\bibitem[Nagashima et~al., 2006]{nagashima2006reproduction}
Nagashima, K., Tsukamoto, K., Satoh, H., Kobatake, H., and Dold, P. (2006).
\newblock Reproduction of chondrules from levitated, hypercooled melts.
\newblock {\em Journal of crystal growth}, 293(1):193--197.

\bibitem[Pack et~al., 2010]{PackEtal:2010}
Pack, A., Kremer, K., Albrecht, N., Simon, K., and Kronz, A. (2010).
\newblock Description of an aerodynamic levitation apparatus with applications
  in earth sciences.
\newblock {\em Geochemical transactions}, 11(1):1--4.

\bibitem[Presnall et~al., 1978]{presnall1978liquidus}
Presnall, D., Dixon, S., Dixon, J., O'donnell, T., Brenner, N., Schrock, R.,
  and Dycus, D. (1978).
\newblock Liquidus phase relations on the join diopside-forsterite-anorthite
  from 1 atm to 20 kbar: their bearing on the generation and crystallization of
  basaltic magma.
\newblock {\em Contributions to Mineralogy and Petrology}, 66(2):203--220.

\bibitem[{Radomsky} and {Hewins}, 1990]{Radomsky&hewins:1990}
{Radomsky}, P.~M. and {Hewins}, R.~H. (1990).
\newblock {Formation conditions of pyroxene-olivine and magnesian olivine
  chondrules}.
\newblock {\em \gca}, 54:3475--3490.

\bibitem[Wasson, 1995]{wasson1995}
Wasson, J. (1995).
\newblock Sampling the asteroid belt: How biases make it difficult to establish
  meteorite-asteroid connections.
\newblock {\em Meteoritics}, 30:595--595.

\bibitem[Wilson et~al., 1999]{WilsonEtal1999}
Wilson, L., Keik, K., Browning, L.~B., Krot, A.~N., and Bourcier, W. (1999).
\newblock Early aqueous alteration, explosive disruption, and reprocessing of
  asteroids.
\newblock {\em Meteoritics and Planetary Science}, 34(4):541--557.

\bibitem[{Zanda}, 2004]{zanda:2004}
{Zanda}, B. (2004).
\newblock {Chondrules}.
\newblock {\em Earth and Planetary Science Letters}, 224:1--17.

\end{thebibliography}

\end{document}